\documentclass{elsart}

\usepackage{amssymb}
\usepackage{latexsym}
\usepackage{amsmath}
\usepackage{amsfonts}

\usepackage{epsfig}

\begin{document}
\journal{European Physical Journal B}
\begin{frontmatter}

\title{Wealth distribution and Pareto's law in the Hungarian
medieval
society}

\author[1,2]{G\'eza  Hegyi},
\author[1,3]{Zolt\'an
N\'eda\corauthref{cor1}}\corauth[cor1]{}
\ead{zneda@phys.ubbcluj.ro},
\and \author[2]{Maria Augusta Santos}

\address[1]{Department of Physics, Babe\c{s}-Bolyai
University,
          str. Kog\u{a}lniceanu 1, RO-400084 Cluj-Napoca,
Romania}

\address[2]{Department of History and Philosophy , Babe\c{s}-
Bolyai University,
          str. Kog\u{a}lniceanu 1, RO-400084 Cluj-Napoca,
Romania}

\address[3]{Departamento de F\'{\i}sica and Centro de
F\'{\i}sica do Porto,
Faculdade de Ci\^encias, Universidade do Porto, Rua de Campo
Alegre 687,
4169-007 Porto, Portugal}

\date{September 2005}

\begin{abstract}
The distribution of wealth in the Hungarian medieval
aristocratic
society is reported and studied. The number of serf families
belonging to a noble is taken as a measure of the
corresponding wealth. Our results reveal the power-law
nature of this distribution function,
confirming the validity of the Pareto law for such a
society. The obtained Pareto index $\alpha=0.92$ is however
smaller
than the values currently reported in the literature. We
argue that the value close to $1$, of the Pareto index is a
consequence of the absence of a relevant economic life in
the targeted society, in agreement with the prediction of
existing wealth distribution models for the idealized case
of independently acting agents. Models developed to explain
city populations may also be adapted to justify our results.

\end{abstract}


\end{frontmatter}

\section{Introduction}

At the end of the XIX century the economist Vilfredo Pareto
\cite{pareto} discovered a universal law
regarding the wealth distribution in societies. His
measurement results on several
European countries, kingdoms and cities for the  XV-XIX
centuries revealed
that the cumulative distribution of wealth (the probability
that the wealth
of an individual is greater than a given value) exhibits a
universal functional form.
Pareto found that in the region containing the richest part
of the population, generally
less than the $5\%$ of the individuals, this distribution is
well described by a power-law
(see for example \cite{ebooks} for a review).
The exponent of this power-law is
denoted by $\alpha$ and named Pareto index. In the limit of
low and medium wealth, the
 shape of the cumulative distribution is fitted by either
an exponential or  a log-normal function.

The power-law revealed by Pareto has been confirmed by many
recent studies on the economy of several corners of the
world. The presently available data is
coming from so apart as Australia \cite{matteo03}, Japan
\cite{aoyama03,fujiwara03}, the US \cite{silva04},
continental
Europe \cite{fujiwara04,clementi04}, India \cite{india} or
the UK \cite{dragulescu01}. The data
is also spanning so long in time as ancient Egypt
\cite{abul02},
Renaissance Europe \cite{souma02} or the $20th$ century
Japan\cite{souma01}. Since it is difficult to measure {\it
wealth}, most of the available data comes from tax
declarations of individual {\it income}
(which is assumed to be proportional to wealth). There are
however some
other databases obtained from different sources, like the
area
of the houses in ancient Egypt \cite{abul02} or the
inheritance taxation
or the capital transfer taxes \cite{UK}. The results mostly
back
Pareto's conjecture on the shape of the wealth distribution.

In the present paper, we present and discuss some recent
results of wealth distribution in a medieval society -- the
Hungarian aristocratic
society around the year 1550. The wealth of the nobles is
estimated and expressed in the number of owned serf
(villein) families, a measure generally used by historians.
The case under study offers a somehow idealized example of a
society without a relevant wealth-exchange mechanism and our
results give further evidence for the universal nature of
Pareto's law.

\section{Statistical Physics approach to Pareto's law}

Typically, the presence of power-law distributions is a hint
for the complexity
underlying a system, and a challenge for statistical
physicists to model
and study the problem. This is why Pareto law is one of the
main problems studied
in Econophysics. Since the value found by Pareto for the
scaling exponent was around $1.5$, Pareto law is sometimes
related
to a generalized form of Zipf's law \cite{zipf} and referred
to as Pareto-Zipf law. According to Zipf's law, many natural
and social phenomena (distribution of words frequency in a
text, population of cities, debit of rivers, users of web
sites, strength of earthquakes, income of companies, etc)
are characterized by a cumulative distribution function
with a power-law tail with a scaling exponent close to $1$.

It is however important to notice that, in contrast to what
happens with most exponents in Statistical Physics, the
Pareto index
$\alpha$, may change from one society to another, and for
the same society can also
change in time depending on the economical
circumstances \cite{fujiwara03,souma01}. The measured values
of $\alpha$ for the  {\it individuals} wealth/income
distribution span a quite broad interval, typically in the
$1.5-2.5$ range; a recent study of
the Indian society \cite{india} reports however, much lower
values ($0.81- 0.92$).
This large variation of $\alpha$ indicates the absence of
universal scaling in this problem -- a feature which models
designed to
describe the wealth distribution in societies should be able
to reproduce.

Models for wealth distribution are essentially defined
by a group of agents, placed on a lattice, that interchange
money following
pre-established rules. The system will eventually reach a
stationary state where some quantities, as for
instance the cumulative distribution of wealth $P_>(w)$,
may be measured. Following
these ideas, Bouchaud and M\'ezard \cite{bouchaud00} and
Solomon and
Richmond \cite{solomon01,biham98} separately proposed a
very
general model for wealth distribution. This model is based
on a
mean field type scenario with interactions of strength $J$
among all the agents
and on the existence of multiplicative fluctuations acting
on each
agent's wealth. The obtained wealth distributions
adjust well to the phenomenological $P_>(w)$ curve. Their
mean-field
results predict that Pareto index $\alpha$ should increase
linearly with
 the strength of interaction between the agents and that
$\alpha=1$ for the case of independent agents ($J=0$).

Similar conclusions were obtained by Scaffeta et al
\cite{scafetta04},
who considered a nonlinear version of the model and from
other regular
lattice based models as those in Refs. \cite{chatterjee03}
and \cite{pianegonda03}.

Given the known complex nature of social networks (see
\cite{dorogovtsev03} for a recent review), models were also
considered in which the economic transactions take place on
a complex small-world or scale-free network
\cite{souma01,iglesias03,garlaschelli04}. More
recently, a model was proposed  \cite{coelho05} where the
network structure is not predefined but rather dynamically
coupled to the wealth-exchange rule. This model successfully
reproduces both the characteristics of the modern family
relation networks and the measured shape of the wealth
distribution curves, accounting for variations in $\alpha$
in the $1.7-2.0$ interval.

From the studies and modeling efforts made by the
statistical physics community, one can conclude that the
emergence of Pareto law can be understood from the
multiplicative nature of the fluctuations of the agents'
wealth and from the dynamics of wealth on the underlying
complex social network.
Both the experimental results and the models suggest that
societies in which there is strong wealth-exchange among the
agents are characterized by higher values of the Pareto
index, whereas societies of isolated agents, where each
agent increases or decreases his wealth in a multiplicative
and uncorrelated manner, are usually characterized by
$\alpha$ close to $1$. Our measurement results on the
Hungarian
medieval society confirm this rule.

\section{Wealth distribution measurements in the Hungarian
medieval society}

To our knowledge, there is no available data concerning the
wealth distribution and the Pareto law for the
Central-Eastern European aristocratic medieval societies. A
flourishing economic life, barter and wealth exchange
developed very slowly in this part of Europe. A centralized
and documented taxation system was also introduced
relatively late. Of course, this undeveloped economic life
makes it hard to collect uniform,
relevant and large-scale data on this society. However, once
such data is obtained, it could be of decisive importance,
since such a society represents an idealized case, where the
agents are acting roughly independently. The underlying
social network and the wealth exchange on it, are expected
to have no major influence, since the aristocratic families
were more or less self-supporting and no relevant barter
existed.
This simplified economic system provides thus an excellent
framework to test, in a trivial case, the prediction of
wealth distribution models.

The first centralized data for 47 districts of the medieval
Hungary (10 of them under Turkish occupation)
is dated from 1550. The data, in a rough format, is
available in a recent book \cite{maksay90}
dealing with property relations of the XVI century Hungary.
For each district, the nobles are arranged alphabetically
and their wealth is grouped in six categories: number of
owned serf families and their lands,
unused lands, poor people living on their land, new lands,
servants and others. Using a method accepted by historians
\cite{engel},
the number of owned serf families is taken as a common
measure of the total wealth of a noble family.

After a careful analysis of all districts and summing up the
wealth for those families
that owned land in several districts, we obtained a dataset
that is usable for wealth distribution studies. We also
imposed a lower cut-off value, chosen to be 10 serf
families, and disregard thus the
low and medium wealth aristocrats. This cut-off is necessary
since
historians suggest that the database is not reliable in this
ranges. With the above constraints, our final database
\cite{dataset} had data for 1283 noble families and 116
religious or city institutions. Considering an average of
five persons per family (a generally accepted value by the
historians and sociologists specialized in the
targeted medieval period) we obtain that our sample contains
around 6400 people. This represents the top $8\%$ of the
estimated $80000$ aristocrats living in Hungary at that
period, and $0.2-0.3\%$  of the estimated total population
(2.7 millions) of Hungary in 1550.

Ranking the considered families after their decreasing
wealth and plotting the rank as a function of wealth for
each family,
gives the tail of the cumulative distribution function for
wealth up to a proportionality
constant (equal to $1/N_t$, where $N_t$
is the total number of families in the
considered society). According to Pareto law, this curve
should have a power-law nature, thus plotting it on a log-
log scale should give a straight line with a negative slope,
which yields the Pareto index. The results obtained from our
dataset are
plotted in Figure 1.

\begin{figure}
\begin{center}
\epsfysize=100mm
\epsffile{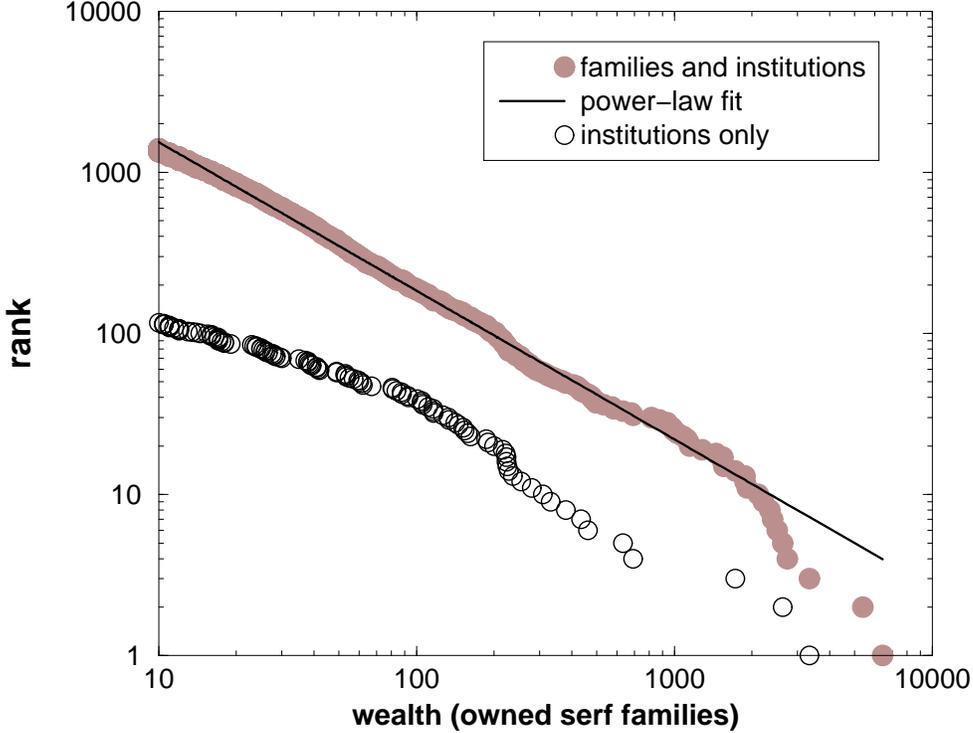}
\caption{The rank of the top $8\%$ aristocrat families and
institutions
as a function of their estimated total wealth
on a log-log scale. Measurement results for the Hungarian
noble society in the year 1550.
The total wealth of a family is estimated as the number of
owned serf families.
The power-law fit suggests a Pareto index $\alpha=0.92$.}
\label{fig1}
\end{center}
\end{figure}

The power-law scaling is nicely visible on two decades in
the figure,  thus confirming Pareto law with an exponent
$\alpha=0.92$. This value is comparable with the ones
obtained for the Indian society \cite {india} and much
smaller than all the other values reported in the
literature.

As explained above, a value of $\alpha$ close to $1$ is in
agreement with our expectations for the studied society.
The validity of the Pareto
law and the value of the Pareto index will not change
considerably (we get $\alpha=0.95$) if we study only
the wealth distribution of the $1283$ aristocratic families
and neglect from the database the $116$ institutions.
Studying however only the wealth distributions of the
institutions will not give a Pareto-like tail at all,
and on the log-log scale we will get a constantly decreasing
slope (empty circles in Figure 1.)
It is also observable from Figure 1, that
the Pareto law breaks down in the limit of the very rich
families, where the wealth is bigger than
1000 serf families. This is presumably a finite-size effect
and such results are observable in other databases
too (\cite{fujiwara03,india}).

In order to have some information on the time evolution of
the Pareto index, the
wealth distribution of the Hungarian aristocratic families
in the 1767-1773 period was also studied.
For this period we had rough data \cite{felho} available
only for 11 districts. The sample was much smaller than for
the year 1550: we had only 531 families and 65 institutions
with total wealth greater than one serf family. However, as
a compensation
for the smaller database, in this case the wealth of each
family is given with three markers: the owned number of serf
families, the exact number of owned serfs and the total size
of the owned land. To our great surprise, we obtained for
each marker
that the cumulative distribution function of wealth does
not give the expected power-law behavior. Instead of a
straight line with a negative slope, we found a constantly
decreasing slope in the log-log plot of the rank as a
function of wealth (Figure 2). The fact that we used only
data for 11 districts cannot be the reason for the breakdown
of the Pareto scaling -- we checked that, on the same 11
districts, the 1550 data still gives the Pareto power-law
tail. We believe the reason why the
Pareto law is not valid for this database, is that a large
wealthy part of the society is missing. Indeed, in the mid
XVIII century in Hungary there were already many wealthy
non-noble families of merchants, bankers, rich peasants,
whose wealth exceeded the wealth of small or middle class
aristocrats. This large category of relatively wealthy
families were not landowners and had no serfs, so they are
simply not present in the considered database. After our
estimates, our wealth
distribution data gives a reliable picture of the society
only for the wealthier aristocrats,
with wealth greater than 100 serf families. It is believed
that the wealth of non-noble families
that owned no land and serfs could not exceed this
threshold.  As observable however in the data from 1550,
for wealth values larger than 1000 serf families finite-size
effects are dominant, and the scaling breaks down.
There is thus a very short wealth interval (one decade)
where the data is trustful. Fitting a power-law
on this interval leads to a Pareto-index $\alpha=0.99$,
which is a reasonable value. This value
is also bigger than the one measured in year 1550,
suggesting thus a more active economic life. From this
Pareto index value it is possible to estimate the wealthy
part of the society
missing from our data (see Appendix A). The results obtained
in such a manner are reasonable ones.

\begin{figure}
\begin{center}
\epsfysize=100mm
\epsffile{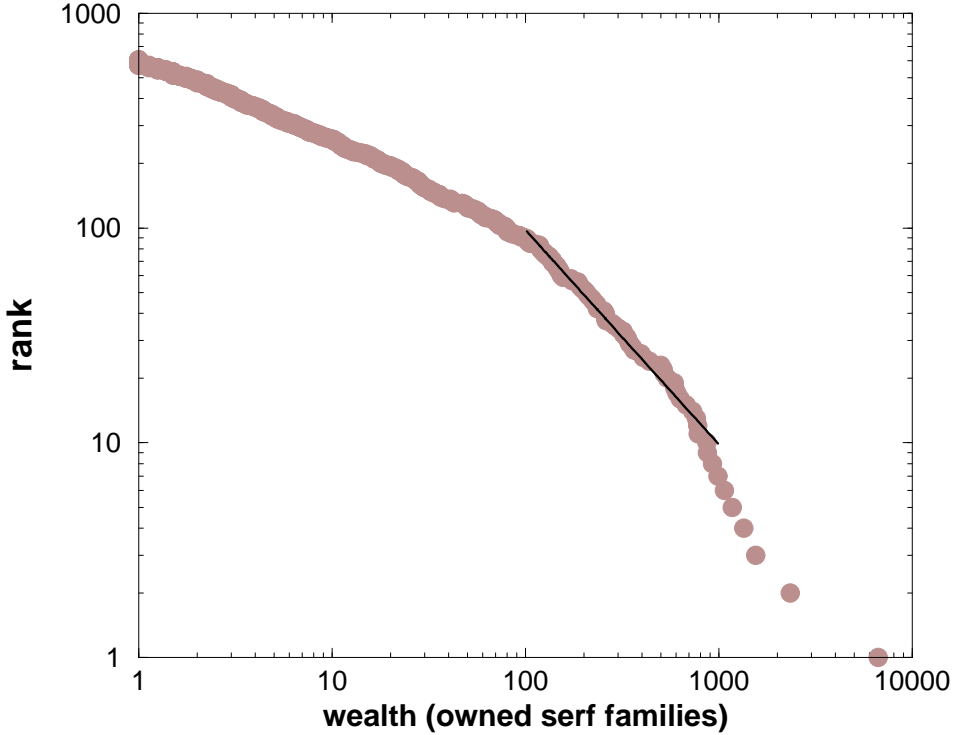}
\caption{The rank of noble families and institutions
as a function of their estimated total wealth
on a log-log scale. Data for the Hungarian noble society
between the years 1767-1773.
The total wealth of a family is estimated in the number of
owned serf families.
The power-law fit suggests a Pareto index $\alpha=0.99$.}
\label{fig2}
\end{center}
\end{figure}

\section{Discussion and conclusions}

Our study shows that the cumulative wealth distribution of
the Hungarian
aristocratic families in the year 1550 exhibits a power-law
shape with a Pareto index $\alpha=0.92$.
This result is a surprise in some sense, since it is
generally believed that
the Pareto index should be bigger than $1$. The fact that
the measured value of $\alpha$ is close
to $1$ is however predicted by existing wealth-exchange
models in the {\it no trade} limit, which we believe is
applicable to the type of society under study.

In the model introduced by Bouchaud and M\'ezard
\cite{bouchaud00},
if we consider the $J=0$ case, the wealth $w_i$ of each
family varies in an independent, stochastic multiplicative
manner, according to

\begin{equation}
\frac{dw_i}{dt}=\eta_i(t) w_i + \xi_i(t)
\end{equation}

where $\eta_i(t)$ is presumably a Gaussian distributed
random variable with zero mean ($<\eta_i>_t=0$), and
$\xi_i$ is an uncorrelated random noise with zero mean
($<\xi_i(t)>_t=0$ and $<\xi_i(t) \xi_i(t')>=C \delta(t-
t')$).
This additive noise is necessary to prevent wealth
extinction and stabilize the power-law type solution in a
finite system. Following
the solution given in \cite{bouchaud00} we immediately get
$p(w) \propto w^{-2}$  for the tail of the wealth
probability density function $p(w)$, which yields a Pareto
index $\alpha=1$ for the cumulative distribution function.
The fact that the measured exponent is smaller than $1$
cannot be explained within this model.

The wealth distribution obtained by us has a shape and
scaling exponent very similar to the population size
distribution of large cities \cite{zipf,zanette}. Zanette
and Manrubia \cite{zanette} successfully justified this
behavior using a simple reaction-like model (which is a
generalization of the  Zeldovich model for intermittency
\cite{zeldovich}).

Inspired by Zanette et al's model, one
can consider thus that the wealth of noble $i$ fluctuates in
time due to several processes, either exogenous (wars,
meteorological conditions affecting harvest sizes,...) or
endogenous (good or bad administration, gambling,...). So
one
can assume  that $w_i (t+1) = \lambda_j w_i (t)$ due to
process $j$, which
occurs with probability $p_j$.  If there are $n$ such
process ( $\sum_{j=1}^n p_j=1$) it is sufficient to require
that $<\lambda>=\sum_{j=1}^n \lambda_j p_j=1$
and to add some noise term to prevent collapse to $0$ to
obtain (see \cite{zanette,comment}) that
$p(w) \propto w^{-2}$ is a stationary solution and thus
$\alpha=1$.
This simple reaction-like stochastic
model is also appropriate thus to understand a Pareto index
close to $1$ for a system composed by independently evolving
agents.

We have thus argued that the obtained Pareto index, close to
$1$, can be justified in the context of several simple
models, compatible with the socio-economical characteristics
of the real system under study. Recalling that a similar
value of $\alpha$ was recently presented for contemporary
Indian society, one may speculate that this is not a mere
coincidence, but rather evidence for the minor role of
exchanges in the dynamics of wealthier Indian people.

The value smaller than $1$
obtained for the Pareto index in the Hungarian medieval
society is however still  a puzzle,
and probably further modeling effort is still necessary to
account for it.

\section{Appendix A}

Let us focus on wealth values greater than one serf family.
If we assume that for wealthy families the cumulative
distribution is scaling as $P_{\ge} (w) \propto w^{-1}$,
the $p(w)$ wealth distribution density function has the
form:

\begin{equation}
p(w)=\frac{C}{w^2}
\end{equation}

The $C$ constant can be determined by taking into account
that we have $83$ families and institutions with wealth
between
$100$ and $1000$ serfs

\begin{equation}
\int_{100}^{1000} \frac{C}{w^2} dw \approx \frac{C}{100}=83,
\end{equation}
leading to $C=8300$. From here it is immediate to estimate
the $N_1$ total number of
families or institutions in the targeted 11 districts of
Hungary that have wealth greater than $1$ serf family:

\begin{equation}
N_1=\int_1^{W_0} \frac{C}{w^2} dw \approx C = 8300
\end{equation}

($W_0$ stands for the biggest reported wealth value,
$W_0=6600$). This estimate can be confronted
with the results of the census made between 1784-1787. For
the studied
11 districts the total population was estimated to be around
1.6 million, with 45 thousand aristocrats
and 33 thousand rich burgers. Taking again $5$ members
per family, our database shows that
only around $531 \times 5=2655$ aristocrats had wealth
bigger than the considered $1$ serf family limit. This
value is only $6\%$ of the total estimated aristocrats in
the society. Most of the aristocrats at that time
had thus no considerable fortune except their noble title.
The missing $8300-600=7700$ families
corresponding to roughly $7700 \times 5=38500$ persons,
should be mostly rich burghers or some richer
peasants. This estimate is in reasonable agreement with the
census from 1784-1787.

{\underline{Acknowledgment}}

Z. Neda acknowledges a Nato Fellowship and the excellent
working atmosphere at the
Centro de F\'{\i}sica do Porto. Research funded by EU
through FEDER/POCTI.

\end{document}